\documentclass[aps,preprint,final,prb,12pt,showpacs]{revtex4}
\usepackage{times}
\usepackage{graphicx}
\begin{document}

\title{
Spin-dependent resonant tunneling through 
quantum-well states in magnetic metallic thin films
\footnote{The submitted manuscript has been authored by a contractor of the
U.S. Government under contract No. DE-AC05-00OR22725.  Accordingly, the
U.S. Government retains a nonexclusive, royalty-free license to publish
or reproduce the published form of this
contribution, or allow others to do so, for U.S. Government
purposes.}
}
\author{
Zhong-Yi Lu and X.-G. Zhang 
}
\affiliation{Computer Science and Mathematics Division,
Oak Ridge National Laboratory, Oak Ridge, TN 37831
}
\author{Sokrates T. Pantelides}
\affiliation{
Dept. of Physics and Astronomy, Vanderbilt University,
Nashville, TN 37235\\
and Oak Ridge National Laboratory, Oak Ridge, TN 37831}

\date{\today}
\begin{abstract}
Quantum-well (QW) states in {\it nonmagnetic} metal layers contained in magnetic multilayers are known
to be important in spin-dependent transport, but the role of QW states in {\it magnetic} layers remains 
elusive. Here we identify the 
conditions and mechanisms for resonant tunneling 
through QW states in magnetic layers and determine candidate structures. We report 
first-principles calculations of spin-dependent transport in
epitaxial Fe/MgO/FeO/Fe/Cr and Co/MgO/Fe/Cr tunnel junctions.
We demonstrate the formation of sharp QW states in the Fe layer and show 
discrete conductance jumps as 
the QW states enter the transport window with increasing bias. At
resonance, the current increases by one to two 
orders of magnitude. The tunneling magnetoresistance ratio is several times larger than in
simple spin tunnel junctions 
and is positive (negative) for majority- (minority-) spin resonances, with a large asymmetry 
between positive and negative biases. The results can serve as the basis for novel spintronic devices.
\end{abstract}
\maketitle
\newpage


Metallic QW states have long been known to play an important role in
spin-dependent transport.
Recently, Yuasa {\it el al}\cite{Science} observed oscillations of the tunneling
magnetoresistance ratio (TMR) in Co/Cu/Al-O/Ni-Fe
tunnel junctions as a function of the thickness of the Cu layer, demonstrating
the effect of quantum confinement of electrons in the nonmagnetic Cu layer.
Most studies, however, have
focused on QW states in a {\it nonmagnetic} layer incorporated in a magnetic multilayer structure. In these
nonmagentic layers, the electron energy dispersion is close to that of a free electron, QW
states can easily be identified, and simple models can be readily applied. There exists
only one report of an experimental attempt to explore spin-dependent transport
through confined {\it magnetic} thin films. In 2002,
Nagahama {\it et al} reported\cite{Nagahama} indications of   
oscillations in the tunneling conductance of a
(100) Cr/Fe/Al-O/FeCo tunneling junction, where QW states were expected to be present
in the magnetic Fe layer. The observed effect was very small,
seen only in the second derivative of the current-voltage curve.
Yuasa {\it el al}\cite{Science} attributed the smallness of the effect to the short mean free
path and spin diffusion length in Fe.  
Resonant tunneling through QW states in magnetic thin films is worth exploring further, 
however. In addition to the intrinsic interest, resonant tunneling through magnetic layers can be switched   
on and off by external magnetic fields, which offers additional control in designing 
spintronics devices.

In this paper, we first discuss several factors that may have contributed to the 
smallness of the effect observed in Ref. \onlinecite{Nagahama}. We then identify 
criteria which, when met by candidate structures containing a magnetic thin film, 
ensure the formation of sharp QW states in the magnetic film and result in 
strong resonant tunneling. We use these criteria to select specific structures 
for a detailed theoretical study. We report results of 
first-principles calculations of spin-dependent transport through select structures 
(bcc (100) Fe/MgO/FeO/Fe/Cr and Co/MgO/Fe/Cr with
epitaxial lattices) and analyse the results. 
We first demonstrate the existence of sharp QW states in the
thin Fe layer and then demonstrate that resonant tunneling occurs through these states
as the bias increases gradually. At resonance, the current 
increases by one to two
orders of magnitude and the tunneling magnetoresistance ratio
(TMR) increases by several orders of magnitude. The resonances are asymmetric with respect to
positive or negative biases. By analysing the wave functions of QW states, we 
demonstrate that the series of resonances
observed in Ref. \onlinecite{Nagahama} as a function of the Fe film thickness 
are from {\it different} QW states that sequentially enter and leave the 
energy window under analysis. 
Finally, we show that a previously unknown QW
state in the Fe minority spin channel contributes to a large and negative
TMR.

We start with a discussion of the experimental data of Ref. \onlinecite{Nagahama}. 
The smallness of the observed effect is likely to be caused by several factors. One factor
is the amorphous nature of the Al-O barrier layer, which results in an 
atomically rough interface (Si/SiO$_2$ is the most abrupt  
known crystal-amorphous interface, but it has roughness of order 1-2
monolayers\cite{SiO}). 
The roughness causes smearing of the energy width of possible QW states,
which reduces resonant-tunnelling effects. A short spin mean free path and diffusion length 
is a possible second factor, as mentioned in Ref. \onlinecite{Science}. 
Indeeed, it has been 
found\cite{Enders} that in thin Fe (100) films grown expitaxially on
a GaAs substrate, the majority spin mean-free-path is about 1.1 nm. The resonance
effects based on majority-spin QW states
would then be limited to Fe films that are thinner than about 1 nm. However,
the minority spin mean free path is much larger, about 14 nm.
If we assume that the spin diffusion length is comparable, 
pronounced QW effects from the minority spin channel should be more
readily observable than from the majority spin channel 
in Fe (100) films of practical thicknesses.
There exists a third factor that can play an important role. 
The energy bands in the magnetic film, the spacer layer, and the 
electrodes, especially their band offsets at the interfaces, 
control the formation of QW states while the Bloch-function symmetries
control the transmission coefficient and hence the magnitude of the resonant tunneling current. 
This point was demonstrated in Refs. \onlinecite{FeMgO, FeOMgO, Mathon, CoMgO},
where the existence of $\Delta_1$ 
Bloch states near the Fermi energy is an essential feature of large TMRs in
FM1/MgO/FM2 
tunnel junctions, where FM1
and FM2 stand for either Fe or Co.
In such systems, distinct differences between the two spin channels, and any
possible sharp resonances arise from the precise matching or mismatching of the
Bloch wave function symmetries on both sides of the barrier. Resonant tunneling is then predicated 
on having a ballistic tunnel junction. From this point of view,
the amorphous nature of the barrier layer may be the critical factor that
surpressed resonant tunneling in the experiments of Ref. \onlinecite{Nagahama}. 

The above analysis leads to two criteria for the selection of candidate structures.
First is the use of a crystalline spacer layer in a structure consisting of crystalline epitaxial 
layers. One obvious choice is MgO, which has already been used in both 
theoretical and
experimental\cite{Yuasa,reports,Parkin} work in simple spin tunnel junctions,
as mentioned earlier. 
Lattice-matching considerations automatically limit the choice of metal layers to the bcc solids of the
3d transition metal series and the orientation of the layers to the (100)
direction.
The second criterion is the suitability of the energy bands and band offsets 
of the thin films and electrodes. The electrode in direct contact with the thin
magnetic film must be nonmagnetic while the
other electrode, separated by the MgO spacer, must be magnetic.
To accentuate the resonance effect through the tunneling process, the QW
states must be produced from a band at the Fermi energy. The symmetry of the
band must
match the complex band structure in the energy gap of MgO so that the QW
states have a higher transmission amplitude through the tunnel barrier than
states with a different symmetry. 
For the bcc solids in the (100) direction with the MgO as
the barrier, the relevant band has
$\Delta_1$ symmetry.
The confinement effect can only be produced if in the nonmagnetic electrode 
the bottom of the $\Delta_1$ band
is well above the Fermi energy.
   
Based on the above criteria, the magnetic film and the magnetic
electrode can be either bcc Fe or bcc Co.
The nonmagnetic electrode must be bcc Cr.
The minority-spin band structure of bcc Fe is very similar to bcc
Cr, while the majority spin has a very different band structure than
Cr.\cite{MJW}
Along the symmetry axis $\Gamma H$, the Fe
majority-spin $\Delta_1$ band crosses the Fermi energy,
but there is no $\Delta_1$ band near
the Fermi energy for either the Fe minority spin nor the Cr band
structure. In addition to the band offsets needed for producing QW states,
these energy-band features provide a differentiation between the two spin channels, which we will
exploit in order to achieve strong spin-dependence in the resonant tunneling
effect.

Using Fe as the magnetic film and the magnetic electrode, our first model system
is the bcc (100) Fe/MgO/FeO/Fe/Cr junction. Here we assume that the Cr
electrode is deposited first, so that the junction contains a
single atomic layer of FeO at the bottom interface between Fe and MgO, which
is observed experimentally\cite{Meyerheim,FeO} and was shown\cite{FeOMgO} to
greatly impact the TMR.  The structure is similar to the tunnel junctions measured in Ref.
\onlinecite{Nagahama}, except for the barrier layer which was Al-O. In our
second model system, we use Co as the magnetic electrode, and assume that
the Co electrode is deposited first, which produces bcc (100) Co/MgO/Fe/Cr.
Unlike the first system,
there has not been clear experimental evidence of an oxide
layer between the Co electrodes and the barrier MgO layer.


Epitaxial ultrathin films are pseudomorphic, which allows us to fix all
bcc Fe,
Co, and Cr lattice constant of 2.86 \AA, corresponding
to that of bulk Fe.
The MgO lattice constant is taken 
to be a factor of $\sqrt{2}$ larger than that of the bulk electrodes,
so that the (100) layers of all the materials can be matched epitaxially.
We assumed no vertical relaxations between the layers.
All calculations were performed using the layer-KKR implementation\cite{LKKR}
of the local-spin density approximation (LSDA) of density functional
theory.
The self-consistent calculations at zero bias were performed in the same
manner as in Refs. \onlinecite{FeMgO, FeOMgO,CoMgO}. 
In all calculations, the thickness of the MgO layer is fixed at eight
atomic layers.
Although the ground state of Cr is antiferromagnetic,
for symplicity of the calculation we assumed that the bulk part of
the Cr electrode is nonmagnetic. The moments of five Cr layers next to the 
QW layer were allowed to relax and they showed small moments with alternating
signs. The self-consistent charge
as a function of bias and the current-voltage curves are calculated
following the procedure in Ref. \onlinecite{Chun}.

\begin{figure}[tbp]
\includegraphics[angle=0,width=0.6\textwidth]{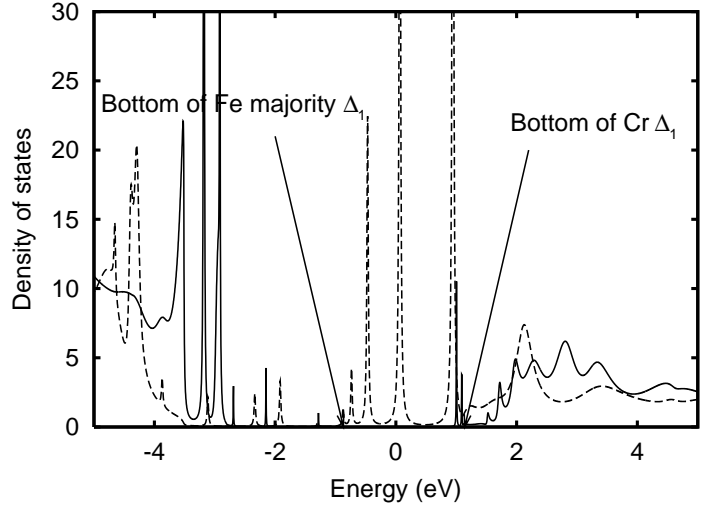}
\caption{$s$-resolved partial density of states
in the eight Fe layers in the
QW film of the Fe/MgO/FeO/8Fe/Cr junction at
the $\bar{\Gamma}$ (${\bf k}_\parallel=0$) point. Solid line: minority spin,
dashed line: majority spin.  The Fermi energy is at 0 eV.}
\label{sdos}
\end{figure}

At the $\bar{\Gamma}$ point, the $\Delta_1$ band is primarily $s$ (angular
momentum $l=0$) but there is
no $s$ component in any of the other bands.
In Fig. \ref{sdos}, 
we show the $s$ partial DOS within the Fe film
sandwiched between the MgO barrier layer and the Cr layer in
bcc (100) Fe/MgO/FeO/8Fe/Cr, with an eight monolayer (ML) thickness of the Fe film.
There is a gap around the Fermi energy with the bottom of the conduction band
coming from the tail of the $\Delta_1$ band in the Cr electrode. Within this
gap, there are several QW states as indicated by the sharp spikes in the DOS.
In the majority-spin channel,
five of these states are above the Fe majority $\Delta_1$ band bottom and are
derived from the majority $\Delta_1$ band.
The width of these peaks equals to exactly twice
the imaginary part of the energy and scales linearly as the imaginary part of
the energy.  This gives us confidence that the
true width of these states is essentially zero, and that they are indeed
QW states and not resonances.

The surprise in Fig. \ref{sdos} is the unexpected QW state for the minority
spin. This is not an interface state since its wave function
extends throughout the Fe film.
The minority spin $\Delta_1$ band bottom of bcc Fe is above the Fermi energy
and is above the $\Delta_1$ bottom of nonmagnetic bcc
Cr at the same lattice constant when the Fermi energy in both solids are
aligned. The possible reason that
a $\Delta_1$ QW state may form here is the
the relative shifts of the atomic potentials on both sides of the interface in
the ultrathin film likely to be different than those of bulk bcc solids.
However, this QW state is not seen in the Co/MgO/9Fe/Cr
junction, indicating that it is probably sensitive to the presence (or absense)
of the FeO layer on the interface.

\begin{figure}[tbp]
\includegraphics[angle=0,width=0.6\textwidth]{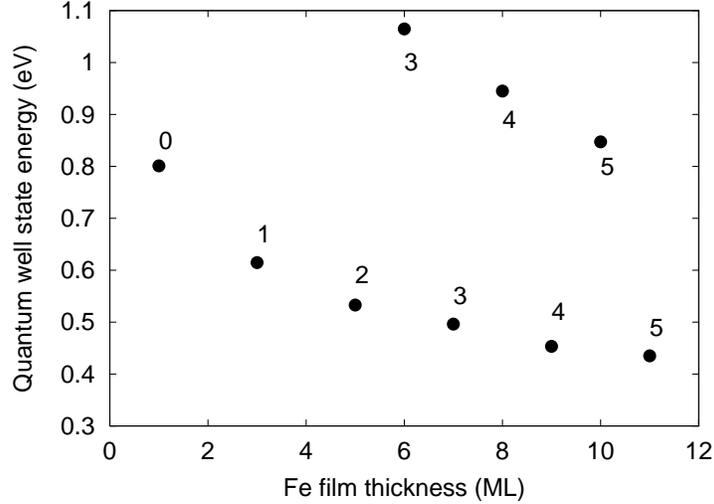}
\caption{The thickness dependence of the majority spin QW state energies for
Fe/MgO/FeO/$n$Fe/Cr for $n=1,\dots,11$. The numbers next to the data points
indicate the number of nodes in the wave function.
}
\label{qw_posi}
\end{figure}
The thickness dependence of the majority spin QW state energies relative to
the Fermi energy is shown in Fig. \ref{qw_posi}
for QW states between 0.3 eV and 1.1 eV. This window is the same as the one in 
which Nagahama et al\cite{Nagahama} observed oscillations in conductance due to the
QW states. If we compare Fig. \ref{qw_posi} with Fig. 2 in Ref. \onlinecite{Nagahama},
the agreement between the two figures is striking. However, the differences
between the two figures are more important because they tell us what we cannot
learn from experimental data alone. We note that in Ref. \onlinecite{Nagahama}
a QW resonance is observed at each monolayer thickness. This observation led to the
assumption that each series of resonances is due to the same QW state. 
The dispersion deduced from this assumption does not fit any known QW states in Fe(100)
films, which led the authors to conjecture that the resonances are due to QW states from
the $\bar{X}$ point rather than the $\bar{\Gamma}$ point. In constrast,
in Fig. \ref{qw_posi}, the lower resonances occur within this energy
window for odd layer thickness only, while the higher resonances occur for
even layer thickness only. Analysis of the wave functions reveals
that each series corresponds to a number of different
QW states, all from the $\bar{\Gamma}$ point. The numbers next to each data
point in Fig. \ref{qw_posi}
indicate the number of nodes of the wave function. This picture tells us that,
when the layer thickness increases by 1 or 2 ML, one QW state moves out of the
energy window and the next QW state moves in.
Ref. \onlinecite{Nagahama} observed QW resonances
at about the same bias voltage for both even and odd layers.
The actual films in that experiment may consist of regions of varying
thicknesses.

\begin{figure}[tbp]
\includegraphics[angle=0,width=0.6\textwidth]{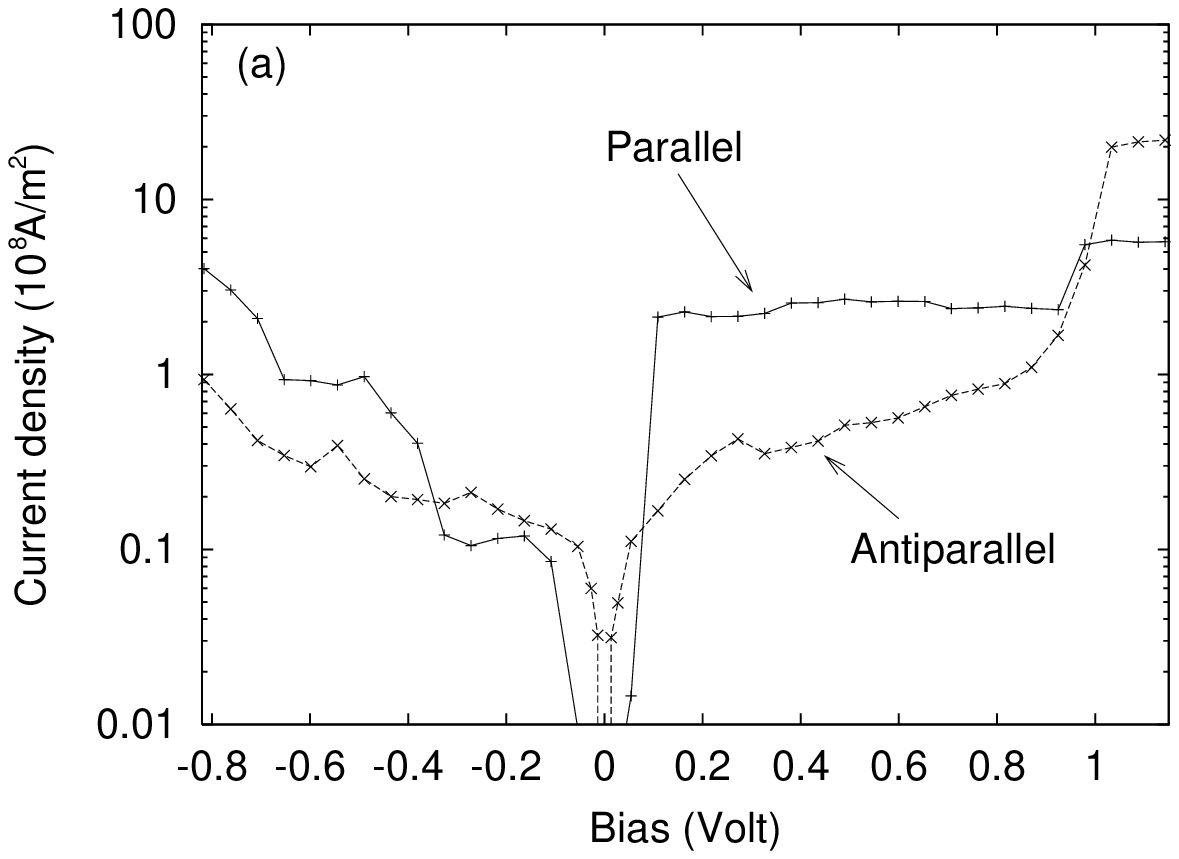}
\includegraphics[angle=0,width=0.6\textwidth]{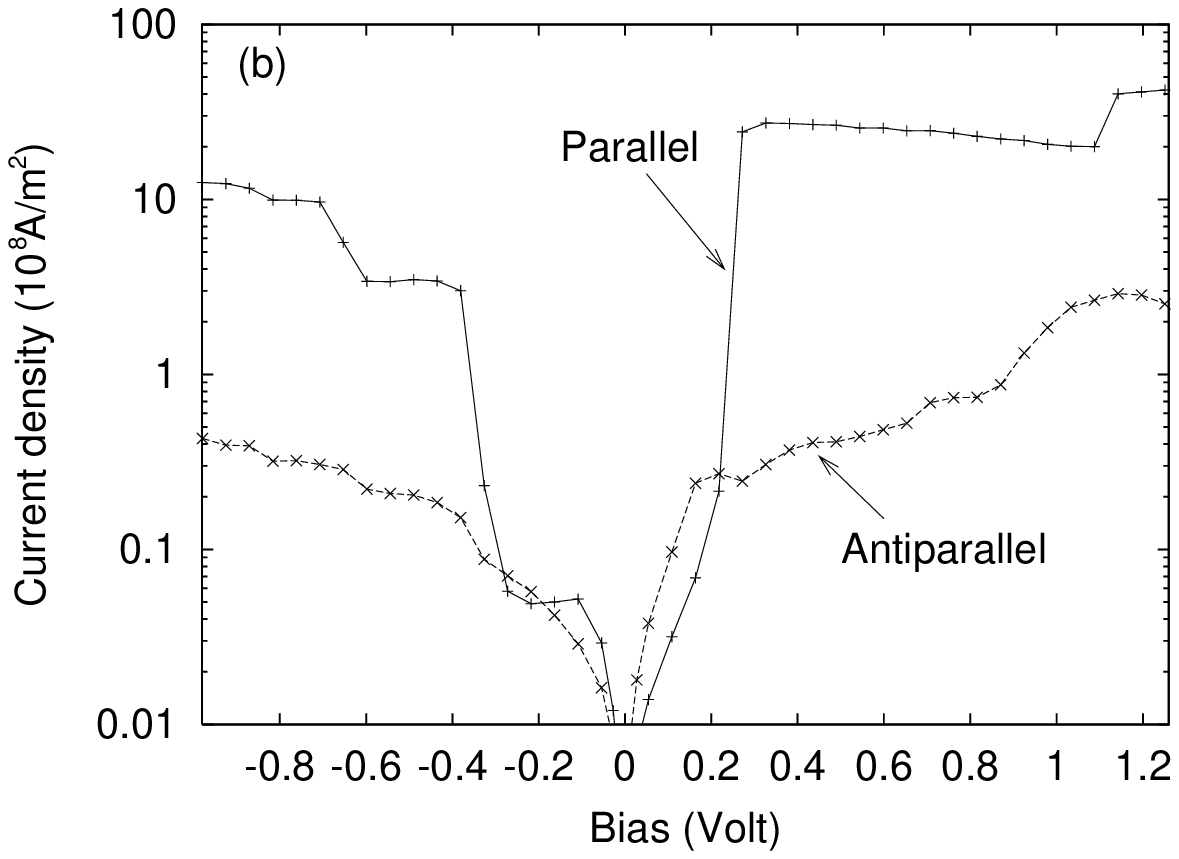}
\caption{Tunneling current as a function of bias voltage. (a) Fe/MgO/FeO/8Fe/Cr,
(b) and Co/MgO/9Fe/Cr. Solid lines are for parallel alignment of the moments
in the Fe film with the magnetic electrode. Dashed lines are for antiparallel
alignment.}
\label{Fe-8Fe}
\end{figure}

In Fig. \ref{Fe-8Fe} we show the calculated I-V curves of the Fe/MgO/FeO/8Fe/Cr,
(a), and Co/MgO/9Fe/Cr, (b), tunnel junctions for both spin channels
with the moment in
the Fe film aligned parallel or
antiparallel to the Fe (or Co) electrode.
A positive bias is defined to make
the Fermi energy of the Cr electrode higher
than that of the ferromagnetic
electrode on the other side of the tunnel barrier. For positive biases,
electrons are injected into the QW states and states above the Fermi
energy are important. Likewise for negative biases the QW states below the
Fermi energy are important.
The tunneling current for the majority-spin channel shows a staircase
feature where the sharp jumps in the current occur at bias voltages that
are precisely equal to the energies of the QW states.
Between these resonances, the current remains nearly
constant. This result indicates that nearly all of the majority-spin current 
flows through the QW states.
The first QW resonance for Fe/MgO/FeO/8Fe/Cr at positive biases
occurs at 
0.068 V. At this voltage, the tunneling current increases by more than an
order of magnitude. The first resonance for Co/MgO/9Fe/Cr is at
0.238 V, where the tunneling current shoots up by two
orders of magnitude. The jumps in the tunneling current for negative
biases at the same voltage values are
smaller.  At low biases, there are no dramatic jumps in the
tunneling current for antiparallel alignment of the moments (the moment of
the Fe film is aligned oppositely with respect to that of the Fe electrode).
Thus, the resonance effect can be easily switched off by applying
a small magnetic field.
For the Fe/MgO/FeO/8Fe/Cr junction, the antiparallel current becomes larger
by almost an order of magnitude
than the parallel current when the positive bias approaches 1 volt. At this voltage,
the QW state in the Fe minority-spin channel enters the transport window, resulting
in a large negative TMR, as we will see below. Such effect is not
seen for the
Co/MgO/9Fe/Cr junction.

\begin{figure}[tbp]
\includegraphics[angle=0,width=0.6\textwidth]{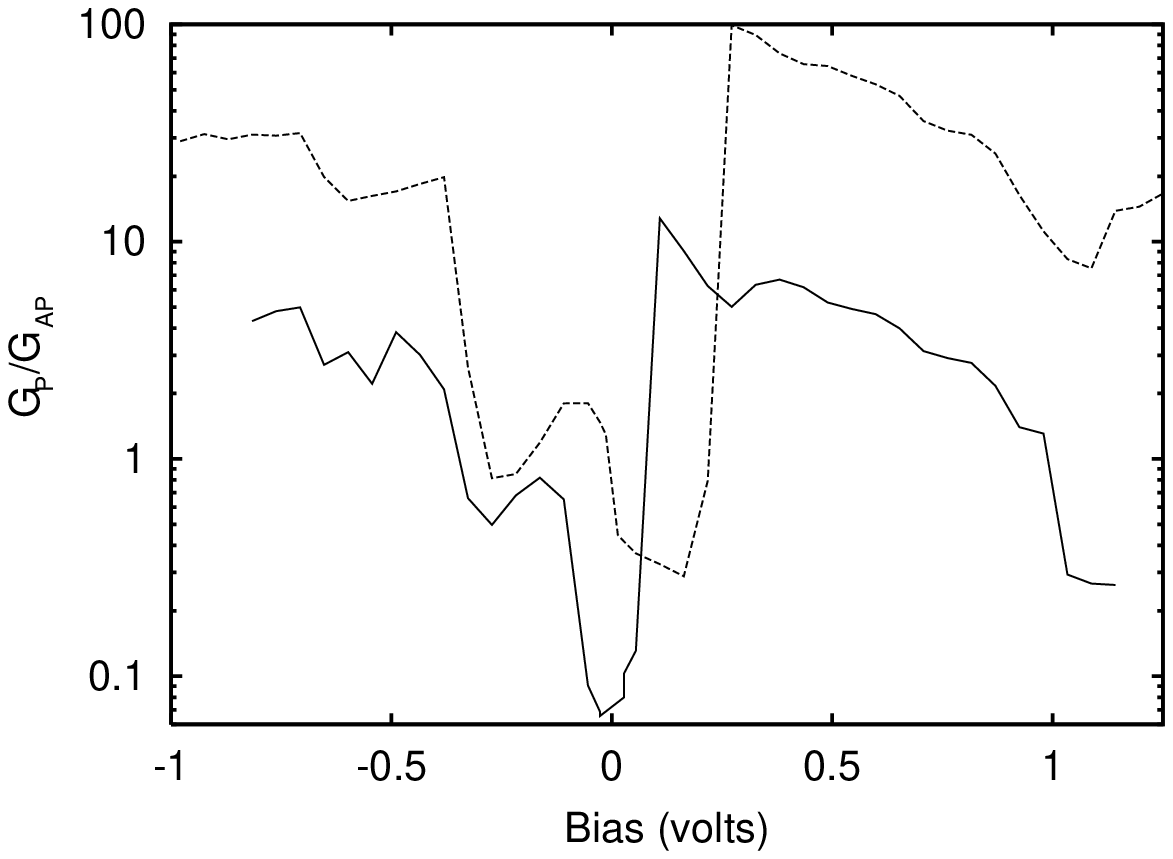}
\caption{Conductance ratio as a function of bias voltage for
Fe/MgO/FeO/8Fe/Cr (solid line) and Co/MgO/9Fe/Cr (dashed line).}
\label{Fe-8Fe-TMR}
\end{figure}

In Fig. \ref{Fe-8Fe-TMR} we show the conductance ratio $G_P/G_{AP}$,
where $G_P$ is the parallel conductance and $G_{AP}$ is the antiparallel
conductance, as a function of the bias voltage
for the same tunnel junctions.
At very small voltages the TMR is sharply negative ($G_P/G_{AP}<1$).
For Fe/MgO/FeO/8Fe/Cr
the ratio $G_{AP}/G_P$ approaches 1600\%
near zero bias. Because one of the
electrodes is Cr, which does not have the $\Delta_1$ state near
the Fermi energy, the majority-spin tunneling current is suppressed at small
biases when the QW states are outside the bias window producing the negative
TMR.  The TMR reaches
peaks at the voltages when each additional QW state starts to
contribute to the tunneling current.  The effect is particularly large for
positive biases. For the Fe/MgO/FeO/8Fe/Cr junction, the first QW
resonance at positive biases produces a maximum TMR of 1200\%.
This value is an order of magnitude larger
than the calculated TMR\cite{Chun} obtained at this voltage for a similar
junction but with Fe electrodes on both sides, whereby the QW states were absent.
For the Co/MgO/9Fe/Cr junction,
the maximum TMR reaches 10000\%. At higher bias, the TMR declines
with voltage because the resonant contribution from each QW state remains
relatively constant while the antiparallel current increases with the voltage.
For the Fe/MgO/FeO/8Fe/Cr junction, the minority spin QW resonance appears near
1 volt and the TMR turns negative again at this bias. Above 1 volt the ratio
$G_{AP}/G_P$ is 380\% .


To summarize, our first-principles calculations predict a significant
QW resonance effect in a class of tunneling junctions including Fe/MgO/FeO/Fe/Cr
and Co/MgO/Fe/Cr.  We predict several-fold increases in tunneling current and
the TMR due to resonant tunneling through the metallic QW states in the Fe
film. Both majority and minority spin QW states can contribute to resonant
tunneling. The former contributes positively to the TMR while the latter
negatively.
We also predict a large asymmetry in the I-V curve between positive and negative biases.
These properties make these junctions excellent candidates for novel
spintronics devices that may combine the features of diodes, rectifiers, 
field effect transistors, and TMR sensors.


This work was supported by the Office of BES
Division of Materials Sciences, and the Office of ASCR
Division of Mathematical, Information and
Computational Sciences of the U.S. DOE.
Oak Ridge National Laboratory is operated by UT-Battelle, LLC, for
the U.S. DOE under contract DE-AC05-00OR22725. The work was further supported
by the DOE grant FDEFG0203ER46096, and by the McMinn Endowment at 
Vanderbilt University.


\begin{references}

\bibitem{Science} S. Yuasa, T. Nagahama, and Y. Suzuki, {\it Science} {\bf 297}, 234 (2002).

\bibitem{Nagahama} T. Nagahama, S. Yuasa, Y. Suzuki, and E. Tamur,
{\it J. Appl. Phys.} {\bf 91}, 7035 (2002).

\bibitem{SiO} A. Bongiorno, A. Pasquarello, M. S. Hybertsen, and
L.C. Feldman, {\it Phys. Rev. Lett} {\bf 90}, 186101 (2003).

\bibitem{Enders} A. Enders, T. L. Monchesky, K. Myrtle, R. Urban, B. Heinrich
J. Kirschner, X.-G. Zhang and W. H. Butler, {\it J. Appl. Phys.} {\bf 89}, 7110 (2001).

\bibitem{FeMgO} W. H. Butler, X.-G. Zhang, T. C. Schulthess, and
J. M. MacLaren, {\it Phys. Rev. B} {\bf 63}, 054416 (2001).

\bibitem{FeOMgO} X.-G. Zhang, W. H. Butler, and Amrit Bandyopadhyay,
{\it Phys. Rev. B} {\bf 68}, 092402 (2003).

\bibitem{CoMgO} X.-G. Zhang and W. H. Butler, {\it Phys. Rev. B}, manuscript number
LQ9658BJ, accepted.

\bibitem{Mathon}
J. Mathon and A. Umerski, {\it Phys. Rev. B} {\bf 63}, 220403(R) (2001).

\bibitem{Yuasa} S. Yuasa, A. Fukushima, T. Nagahama, K. Ando, and
Y. Suzuki, {\it Jap. J. Appl. Phys.} {\bf 43}, L588 (2004).

\bibitem{reports}
X. Jiang, {\it Bulletin of APS}, {\bf 49} 1079, (2003). R. Wang, X. Jiang, R. Shelby,
R. MacFarlane, S. Bank, J. Harris, S. Parkin, {\it Bulletin of APS} {\bf 49} 1079 (200
3).
\bibitem{Parkin}
S. S.P. Parkin, C. Kaiser, A. Panchula, P. Rice, B. Hughes, M.
Samant, and S.-H. Yang, preprint.

\bibitem{Meyerheim} H. L. Meyerheim, R. Popescu, J. Kirschner, N. Jedrecy,
M. Sauvage-Simkin, B. Heinrich, and R. Pinchaux, {\it Phys. Rev. Lett.} {\bf 87},
076102 (2001).
\bibitem{FeO}
H. L. Meyerheim, R. Popescu, N. Jedrecy, M. Vedpathak, M. Sauvage-Simkin,
R. Pinchaux, B. Heinrich, and J. Kirschner, {\it Phys. Rev. B} {\bf 65}, 144433 (2002).

\bibitem{MJW} V. L. Moruzzi, J. F. Janak, and A. R. Williams,
{\it Calculated Electronic Properties of Metals}, (Pergamon, New York, 1978).

\bibitem{Chun} C. Zhang, X.-G. Zhang, P. S. Krsti\'{c}, H.-P. Cheng,
W. H. Butler, and J. M. MacLaren, {\it Phys. Rev. B} {\bf 69}, 134406 (2004).

\bibitem{LKKR} J. M. MacLaren, X.-G. Zhang, W. H. Butler, and Xindong Wang,
{\it Phys. Rev. B} {\bf 59}, 5470 (1999).
\end{references}
\end{document}